\documentclass[fleqn,twoside]{article}
\usepackage{espcrc2}
\usepackage{epsf,amssymb}
\usepackage[dvips]{graphicx}

\newcommand{\cd}{\!\cdot\!}
\newcommand{\td}{\mathrm d}

\newcommand{\AmS}{{\protect\the\textfont2
  A\kern-.1667em\lower.5ex\hbox{M}\kern-.125emS}}

\title{
{ \vspace{-30mm} \normalsize \hfill
\parbox{30mm}{COLO-HEP-477\\October 2001}
}\\[23mm]
 The Static Potential with Hypercubic Blocking}

\author{A. Hasenfratz\address[CU]{Department of Physics, University of Colorado,
        Campus Box 390,   Boulder, CO. 80309, USA},
        R. Hoffmann\addressmark[CU]\thanks{currently at Universit\"at Regensburg}
        and
        F. Knechtli\addressmark[CU]\thanks{currently at Humboldt Universit\"at
        Berlin}}

\begin{document}

\begin{abstract}
We measure the static potential from Wilson loops constructed
using
 hypercubic blocked (HYP) links. The HYP potential agrees
 with the potential measured using thin links for distances $r/a\geq2$. We
 calculated the lowest order perturbative expansion of the lattice Coulomb
 potential of HYP links. These results are used in analyzing the static
 potential both on quenched and dynamical lattices. The statistical
 accuracy of the potential with HYP links improves by about an order of
 magnitude, giving a reliable scale even
 with limited statistics both on
quenched and dynamical lattices. \vspace{-0.8pc}
\end{abstract}

\maketitle

\section{Introduction}
Recently \cite{hyp} Hasenfratz and Knechtli introduced the
hypercubic blocking transformation (HYP) and used it to study
dynamical staggered fermions with greatly improved flavor
symmetry. Here we use the HYP blocking as an operator to measure
the static potential from Wilson loops. The HYP smearing mixes
gauge links within hypercubes attached to the original link only,
therefore it has less impact on the short-distance properties of
the gauge configurations than repeated levels of APE blocking. The
statistical accuracy of the potential measured with HYP links
improves by about \emph{an order of magnitude}.

We calculated the lowest order perturbative expansion of the
lattice Coulomb potential of HYP links. The distortion effect of
the smearing is visible at distances $r/a\lesssim2$ but is
negligible beyond that. By removing lattice artifacts using the
perturbative Coulomb potential we could fit the static potential
with the continuum form even from distance $r/a\!\approx\!1$ with
low $\chi^2$. This, combined with the improved statistical
accuracy at large distances makes it possible to get reliable
values for both $r_0$ and $\sqrt\sigma$ even with limited
statistics.

\section{Constructing the HYP link}

A single level of HYP smearing consists of three levels of
modified APE smearing, where the staples are restricted in such a
way that only links within hypercubes attached to the original
link are used. The construction equations can be found in
\cite{hyp}. There are three free APE-blocking parameters
$\alpha_i$ ($i=1\ldots3$). These were optimized
\emph{non-perturbatively} to suppress fluctuations at the
plaquette level. The \emph{perturbative} cancellation of the
flavor symmetry violating terms at tree level of the staggered HYP
action gives consistent values for the
  blocking parameters \cite{hafra}.

The figure shows a schematic representation of the construction in
3D: The fat link is built from the four double-lined staples
(there are six in 4D) (a), which \hspace*{0.05pt} in turn \hspace*{0.05pt} are
constructed \hspace*{0.1pt} from the
two\\[-6.1mm]

\begin{figure}[!h]
\begin{center}
\parbox[!h]{27mm}{\vspace*{-28.5mm} staples (four in 4D) that
stay  within  the hypercubes attached
to the original
link (b). The last step in the construction
 is not doable in three dimensions.}
\hspace*{0.0cm} \includegraphics[height=30mm]{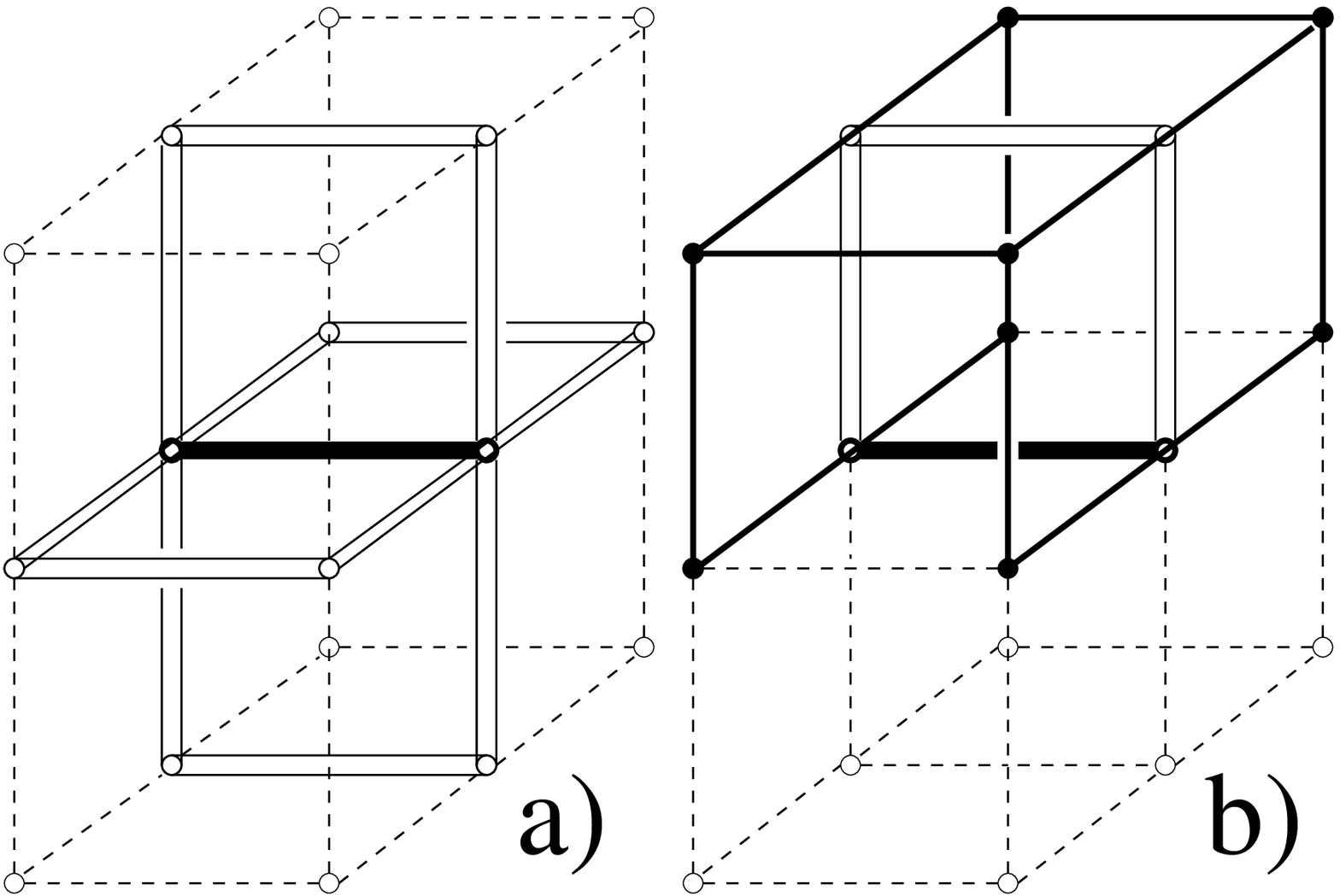}
\end{center}
\end{figure}

\vspace*{-12.4mm}

\section{Perturbative potential}

Let the gauge fields for \vspace*{1.5pt} the thin links be $A_\mu(x)$ and those
for the HYP \vspace*{1.5pt} links $B_\mu(x)$. In Fourier space they are related
as: {\small
\begin{eqnarray*}
B_\mu(p)&\hspace*{-4mm}=\hspace*{-4mm}&\sum_\nu h_{\mu\nu}(p)A_\nu(p)+\mathcal O(A^2),\ \ \textrm{where }\\
 h_{\mu\nu}(p)&\hspace*{-4mm}=\hspace*{-4mm}&\delta_{\mu\nu}\Big[1\!-\!\frac{\alpha_1}6
\sum_\rho \hat
p_\rho^2\Omega_{\mu\rho}(p)\Big]\!+\!\frac{\alpha_1}6\hat p_\mu
\hat p_\nu\Omega_{\mu\nu}(p),\\
\Omega_{\mu\nu}(p)&\hspace*{-4mm}=\hspace*{-4mm}&1+\alpha_2(1\!+\!\alpha_3)-\frac{\alpha_2}4
(1\!+\!2\alpha_3)(\hat p^2\!-\!\hat p_\mu^2\!-\!\hat
p_\nu^2)\\&\hspace*{-3mm}&+\frac{\alpha_2\alpha_3}4{\prod_{\eta\neq\mu,\nu} \hat p_\eta^2}\\
\textrm{and}\ \hat p_\mu&\hspace*{-4mm}=\hspace*{-4mm}&2
\sin(p_\mu/2)
\end{eqnarray*}}
Using this relation one can calculate the Coulomb part of the
static quark potential measured with HYP links in lowest order
perturbation theory. For a Wilson loop made from HYP
links with a spatial extension $a\cd\mathbf n$ the potential is:\\[-10mm]
\begin{center}{\small
$$ a\cd V_{pert}(a\cd\mathbf n)=\displaystyle-\frac{4}{3}\, g^2\!\int_{-\pi}^{\pi}\frac{\td^3p}{(2\pi)^3}\ \frac{\cos(\mathbf n\cd\mathbf p)
\times\big[\!\!\!\begin{array}{c}
\textrm{\scriptsize smearing}\\[-1.6mm]
\textrm{\scriptsize factor}
\end{array}\!\!\!\big]
}{\sum_{i=1}^3\hat p_i^2} $$}

\end{center}\vspace*{-6mm}
\normalsize

\noindent where: [{\small smearing factor}]=

\vspace*{-3mm}

\begin{eqnarray*}
\quad\left\{\!\!\!\!
\begin{array}{cl}
\left[1-\frac{\alpha_1}6\sum_{i=1}^3 \hat p_i^2\Omega_{i0}\right]^2& \textrm{\small for the HYP potential}\\[2mm]
1& \textrm{\small for the thin potential}
\end{array}\right.\\[2mm]
\end{eqnarray*}

\vspace*{-4mm}

Here we use the Wilson plaquette gauge action. The generalization
to other gauge actions modifies the
propagator in the above formula.\\[-6mm]

\normalsize

\section{Potential measurements}
\linespread{1.06} \small\normalsize The improvement in statistics
due to the HYP smearing is best seen by comparing the static quark
potential measured from Wilson loops using thin and HYP links on
the \emph{same configurations}. The set consists of 240
$8^3\times24$ configurations generated with the Wilson plaquette
action at $\beta=5.7$. The thin link potential has been shifted by
an unphysical constant to match
the HYP potential for $r/a>2.5$:\\[-2mm]
\begin{center}
\includegraphics[width=74mm]{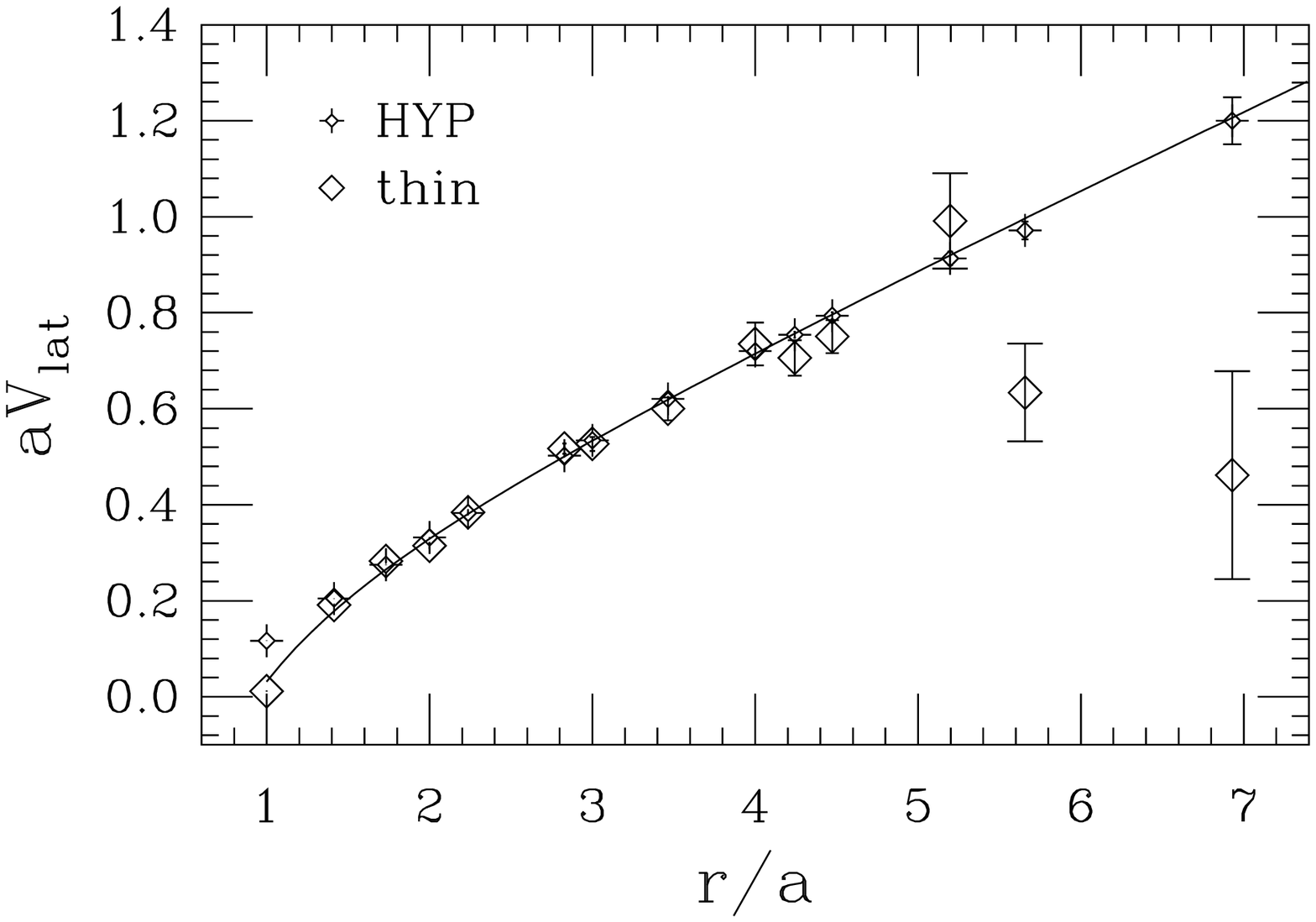}

\end{center}

For distances $r/a\gtrsim\sqrt2$ the HYP and thin potential agree
well and for larger distances the statistical error in the HYP
potential is extremely reduced by the smearing transformation.
This is especially important for a precise measurement of the
string tension. The solid line
is a perturbatively improved fit that will be discussed in the following section.\\[-6mm]

\section{Improved fit}
The deviation of $V_{pert}$ from the continuum Coulomb potential
$V_C(r)=-\frac{g^2}{3\pi r}$
is best seen in an $r$ vs. $r\cd V$ plot:\\[-1mm]

\begin{center}
\includegraphics[width=75mm]{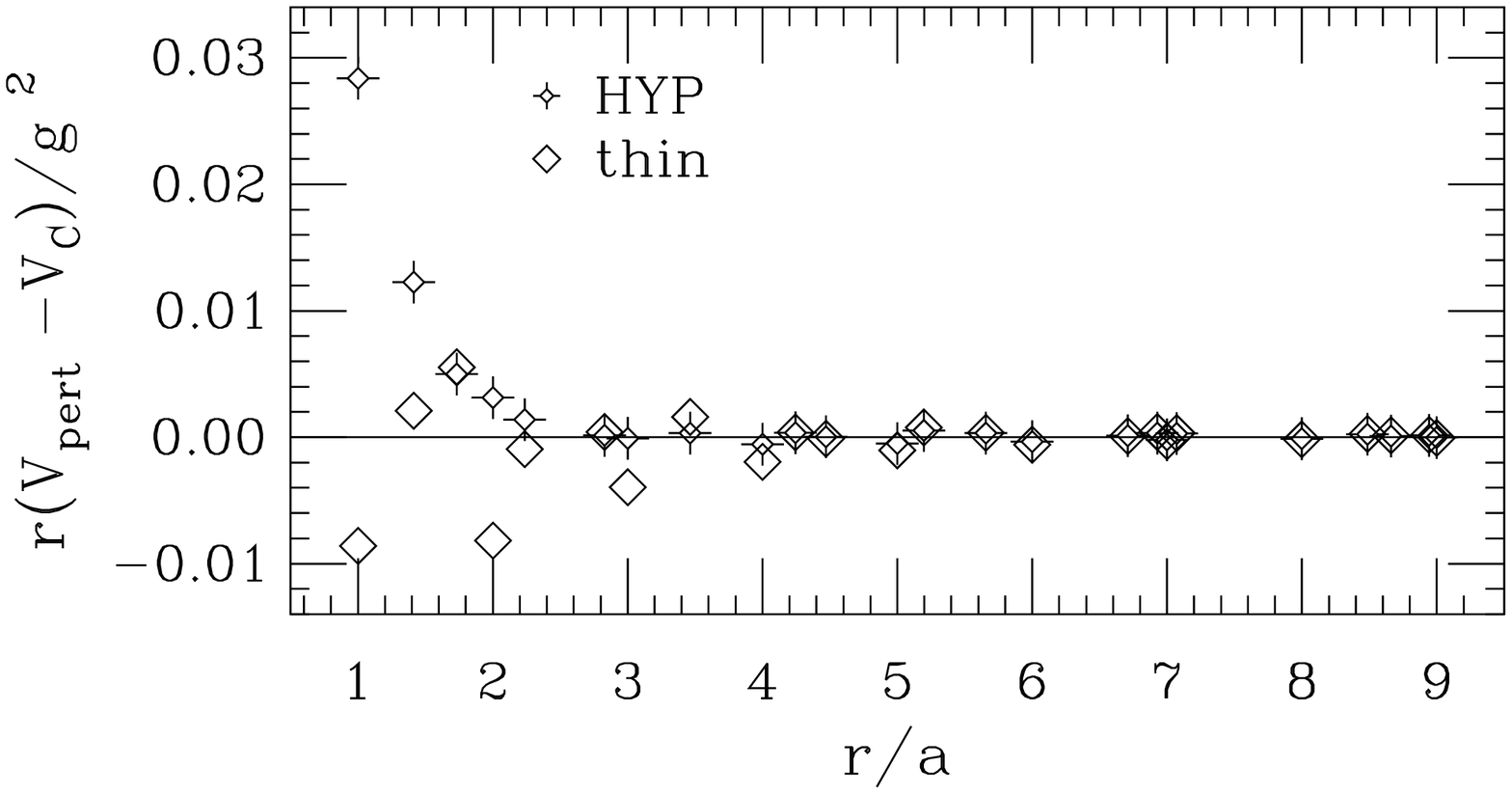}
\end{center}
\vspace*{0mm}

The perturbative calculation captures the short distance
distortion of the HYP smearing and the improved rotational
symmetry for larger distances.

With this perturbative data one can make a four parameter fit to
the lattice data. If we assume that most of the lattice artifacts
are captured by $V_{pert}$ we can write $V_{lat}$ as

\vspace*{-5mm}

\begin{eqnarray*}
V_{lat}&=&V_{cont}+\Delta V_{lat}\ \ \textrm{\small where}\\[1mm]
V_{cont}(r)&=&\frac{c_0}r+c_1+c_2r\\
&&\textrm{\small is the continuum potential and}\\[1mm]
\Delta V_{lat}(r)&=&\widetilde c_0\left(V_{pert}(r)\!-\!V_C(r)\right)\\
&&\textrm{\small represents the lattice artifacts.}
\end{eqnarray*}

\vspace*{-2mm}

$\widetilde c_0,c_0,c_1$ and $c_2$ are the four parameters of the
improved fit and from $c_0$ and $c_2$ the string tension $\sigma$
and Sommer scale $r_0$ can be obtained in the usual way.
\vspace*{-2mm}

\section{Results from the improved fit}
For both thin and HYP potentials the following plot shows the
actual potential data and the ''corrected'' data, where the lowest
order lattice correction is removed, namely: $V_{lat}-\Delta
V_{lat}$. The continuum part of the fit, $V_{cont}$, is shown as a
solid line. The fit uses all data points in the range
$r/a\in(0.9,5)$:\\[-5mm]

\begin{center}
\includegraphics[width=74mm]{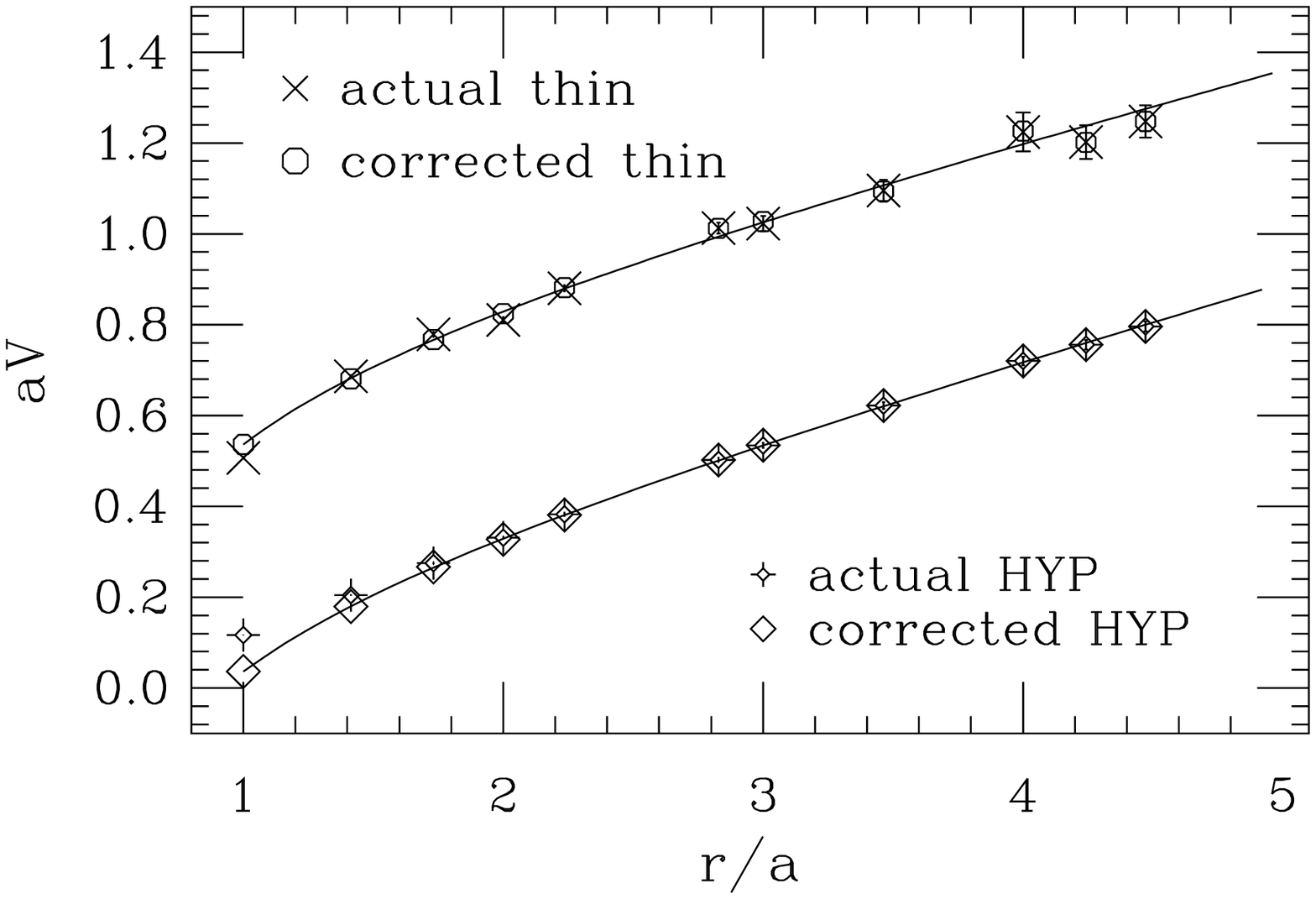}
\end{center}
\vspace*{-4mm}

After removing the lattice artifacts and adjusting an overall
constant the thin and HYP measurements match very closely.

The perturbatively corrected potential shows very good rotational
symmetry and closely follows the continuum form over the whole
range of $r/a$. The $\chi^2$ per degree of freedom decreases by
about a factor of 10 if we include the perturbative correction in
the fit. The correction of the lattice artifacts is so effective
that one can use all the points, even $r/a=1$, for fitting. As an
example the following plot shows the dependence of $\sqrt\sigma$
on the lower limit of the fit ($r_{min}$): \vspace*{-2mm}

\begin{center}
\includegraphics[width=74mm,clip=true]{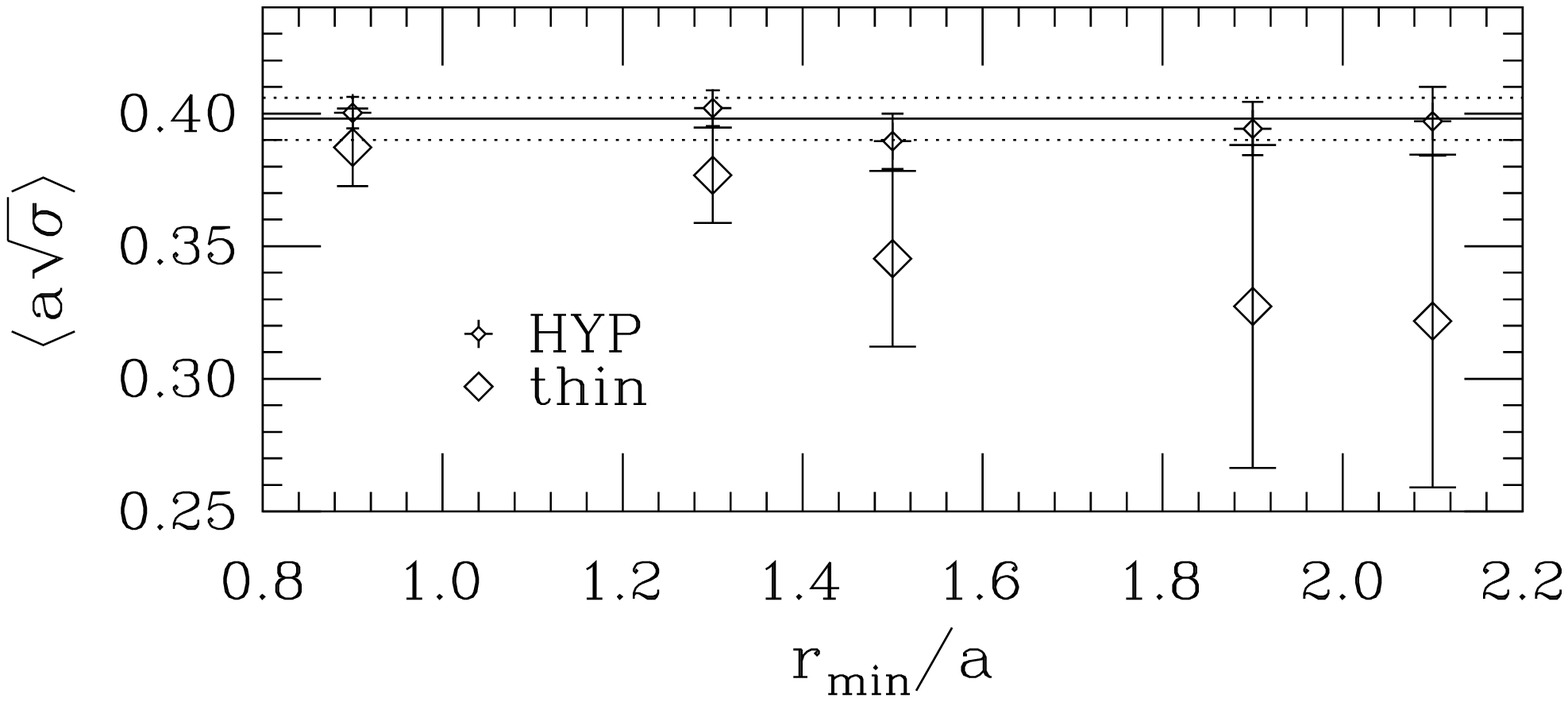}
\end{center}

\vspace*{-3mm} The results for the potential parameters of the
$\beta=5.7$ quenched lattices from standard jackknife analysis are
(together with data from some large scale
simulations at the same coupling value):\\[-2mm]
\small
\begin{center}
\def\arraystretch{1.2}
\begin{tabular}{|l|l|c|c|}\hline
\multicolumn{1}{|c|}{$\mathbf{r_0/a}$}   &
\multicolumn{1}{|c|}{$\mathbf{a\sqrt\sigma}$}   &   $\!N_{conf}\!$
&type of link\\\hline $\ \mathbf{2.93(2)}$   & $\mathbf{0.398(8)}\
$    &   240     & {\textrm{HYP}}\\\hline $\ 2.922(9)$ & $\qquad$-
& 1000        & {\small multi-hit \cite{alpha}}\\\hline $\
2.990(24)$  &   $0.3879(34)$       & 4000 & {\small thin
\cite{edwards}}\\\hline
\end{tabular}
\end{center}

\vspace*{-1mm}

\normalsize

\section{Summary}
We have measured the static potential using HYP smeared links.\\

\hspace*{-10mm}\begin{minipage}{7.6cm}
\begin{itemize}
\item  The HYP potential has greatly reduced statistical errors and shows improved rotational symmetry.\\[-6mm]
\item  The short distance potential is distorted\\ only at $r/a<2$.\\[-6mm]
\item  The perturbative HYP Coulomb potential describes most of these distortions.\\[-6mm]
\item With the perturbative correction  potential data even at $r/a<2$ can be used for
fitting.\\[-1mm]
\end{itemize}
\end{minipage}

\noindent The improved statistical accuracy and the perturbative
correction make the HYP potential measurement very effective for
the analysis of both quenched and dynamical configurations. We
would like to emphasize that at present other refined techniques,
like multi-hit \cite{mh} or multi-level \cite{ml}, are not
applicable with dynamical fermions in contrast to the HYP
smearing, with which we were able to obtain reliable scale
determinations on dynamical lattices even with limited statistics.

\end{document}